\documentclass[12pt]{article}

\usepackage{amsmath,amssymb,graphicx} 

\usepackage{epsf}
\usepackage{cite}

\def\beq{\begin{equation}}
\def\eeq{\end{equation}}
\def\bea{\begin{eqnarray}}
\def\eea{\end{eqnarray}}

\def\centeron#1#2{{\setbox0=\hbox{#1}\setbox1=\hbox{#2}\ifdim
\wd1>\wd0\kern.5\wd1\kern-.5\wd0\fi
\copy0\kern-.5\wd0\kern-.5\wd1\copy1\ifdim\wd0>\wd1
\kern.5\wd0\kern-.5\wd1\fi}}
\def\ltap{\;\centeron{\raise.35ex\hbox{$<$}}{\lower.65ex\hbox{$\sim$}}\;}
\def\gtap{\;\centeron{\raise.35ex\hbox{$>$}}{\lower.65ex\hbox{$\sim$}}\;}

\def\singlespaced{\baselineskip=\normalbaselineskip}

\def\dslash{\not{\hbox{\kern-2pt $\partial$}}}
\def\Dslash{\not{\hbox{\kern-4pt $D$}}}
\def\Oslash{\not{\hbox{\kern-4pt $O$}}}
\def\Qslash{\not{\hbox{\kern-4pt $Q$}}}
\def\pslash{\not{\hbox{\kern-2.3pt $p$}}}
\def\kslash{\not{\hbox{\kern-2.3pt $k$}}}
\def\lslash{\not{\hbox{\kern-2.3pt $l$}}}
\def\qslash{\not{\hbox{\kern-2.3pt $q$}}}
\def\epsilonslash{\not{\hbox{\kern-2.3pt $\epsilon$}}}

\newcommand{\newc}{\newcommand}
\newc{\qbar}{{\overline q}}
\newc{\Kahler}{K\"ahler }
\newc{\deltaGS}{\delta_{\rm GS}}

\begin{document}
\begin{titlepage}
\begin{flushright}
{\tt hep-th/0403074}\\
CALT-68-2478

\end{flushright}


\begin{center}
{\huge \bf Gravitational Perturbations\\
\vspace{2mm} of a Six Dimensional Self-Tuning Model}
\end{center}


\begin{center}
{\bf M.L. Graesser,
 J. E. Kile,
 P. Wang}
\end{center}

\begin{center}
{\it California Institute of Technology, 452-48, Pasadena,
CA
91125}
\end{center}


\begin{abstract}
\noindent We investigate gravitational perturbations in
a compact
six-dimensional self-tuning brane model.  We specifically
look
for analytic solutions to the perturbed Einstein equations that
correspond in four-dimensions
to massless or approximately massless scalars
coupled to matter
on the brane. The presence of such modes
with gravitational couplings would be phenomenologically
unacceptable. The most general solution
for all such modes is obtained, but it is
found that they are all eliminated
by the boundary conditions. Our main result is
that to linear order in
perturbation theory this model does not
contain any light scalars. 
We speculate
that this model does not self-tune.

\end{abstract}

\end{titlepage}

\newpage

\setcounter{footnote}{0} \setcounter{page}{2}
\setcounter{section}{0} \setcounter{subsection}{0}
\setcounter{subsubsection}{0}


\section{Introduction}

There has been much speculation on the possibility that the
Standard Model fields are confined on a four-dimensional brane in
a higher-dimensional universe \cite{brane}. The
usual cosmological constant
problem is reformulated in these theories, since in general
the cosmological constant in the four dimensional
effective theory receives contributions from both
bulk physics and from brane physics.
The cosmological problem in these models is balancing
the bulk terms against the vacuum energy on the brane
to
produce the very small value seen in
nature.

A general class of five-dimensional models \cite{selftuning1}
were introduced to partially resolve this problem.
Instead of canceling bulk terms against
brane terms, these models
have the interesting feature
that flat space solutions always exist for arbitrary
values of the tension on the brane.
This is a big improvement, since for these class of
models there is a
hope that the cosmological
constant problem could be immune
to phase transitions on the brane, although this
has not been demonstrated.


It has been recognized
that six-dimensional brane world
models might be more promising
for the cosmological constant problem, due to
the co-dimension-two nature of the geometry \cite{rubakov,sundrum1,luty}.
As might be expected by locality,
a three brane with arbitrary tension
in six-dimensional flat space does
not cause the space to inflate,
but just introduces a conical singularity.
But as such a model does not lead to four-dimensional
Einstein gravity, one must compactify
the transverse space.

An interesting co-dimension-two spherical
compactification was originally considered
by Sundrum \cite{sundrum1}, and modified by
Carroll and Guica \cite{car} and also
by Navarro \cite{navarro}. In their model
a fine-tuning of bulk parameters is required to
obtain a small four dimensional cosmological
constant. What is intriguing though, is that
this fine tuning
is independent of the tension on the brane.
With non-vanishing brane tension the bulk
geometry is still locally a sphere, but
globally it has a conical deficit angle.
Pictorially, a ``banana peel'' has been
removed.

A generic difficulty with
the five-dimensional models was that
self-tuning required a curvature
singularity in the bulk \cite{selftuning2}.
Resolving the singularity by introducing another
brane in the bulk reintroduces a fine-tuning of
the brane tensions \cite{selftuning3}.
Naively the six--dimensional model may be an improvement on the
five-dimensional self-tuning models, since
here the curvature singularities are conical and
perhaps less severe.
Still, this may be too much.
\cite{gregory} finds that in geometries with conical
singularities it is inconsistent to add anything
other than tension to the brane. Since we
would like to also have stars and dogs on the brane,
this is clearly problematic. It is unclear
though whether this is an artifact of treating
the brane as infinitely thin \footnote{Finite thickness 
co-dimension--2 models have been discussed in 
\cite{giovani}.}, and it would
be interesting to see whether higher
dimension operators on the brane could
overcome this obstacle\footnote{\cite{gregory} did find
that adding the Gauss-Bonnet
operator in the bulk
could lead to Einstein gravity with arbitrary sources.}.

In this note we will not dispel this concern.
Rather, we address an independent issue, which is
whether this six-dimensional model leads
to a scalar-tensor theory of gravity at low
energies. Given the presupposition of a
self-tuning mechanism, one might expect such a mode is necessary, 
for example, to self-tune a change in the brane tension 
that is much smaller than the compactification scale.

As is well known,
gravitational couplings of a scalar admixture to
gravity are strongly constrained by measurements \cite{pdg}
of the Nordtvedt effect \cite{nordtvedt}. Other
phenomenological constraints could be obtained
from cosmology, for here
parameters in the four-dimensional
effective theory depend on the brane tension,
which probably had
a cosmological history \cite{nilles}.

Our main result is that
while we do find massless scalars allowed
by the bulk equations of motion, these are all
eliminated by the boundary conditions.
Our analysis can also be extended
to exclude light scalars
whose mass vanishes in the limit that the
tension goes to zero.
To linear
order in perturbation theory then, this model
does not have any phenomenological difficulties
of this sort. Our results support the conclusions of
\cite{cline}, but we improve on their work since here
we are able to obtain all of our results
analytically, without having to resort to a
numerical analysis.

If this model does
have a self-tuning mechanism, then
the absence of any massless scalars does
raise a puzzle 
though, for there is no light scalar 
to adjust a change in the brane tension. 
If this model does in fact have a self-tuning mechanism, 
then our results  
%
surprisingly suggest that by default, 
it is the collective motion
of many massive Kaluza-Klein states that is  responsible for
canceling a change in
the tension.

But there is another reason to doubt whether there model has a
self-tuning mechanism.
For here the deficit angle
is an integration parameter that may be
chosen to satisfy the boundary conditions after 
assuming a four-dimensional flat space ansatz. However,
one might be worried that this feature may not be sufficient 
in order realize the self-tuning. We elaborate on
this in Section 4.


The content of this paper is as follows. In section 2, we briefly
review the key features of the model.  In
section 3.1, we determine
Einstein's equations to linear order
and discuss the appropriate
gauge fixing. Here there is a subtlety
due to a brane bending mode, and we discuss
how we gauge fix this mode. Then we present our solutions
for the most general massless scalar modes
that could couple to matter on the brane. In
section 3.2 we discuss the boundary conditions.
After imposing these conditions, none of our
zero modes survive.
Section 4 reconsiders the self-tuning feature of this model.
Section 5 contains some concluding remarks.

A few notes on notation are in order.  We will use
$G_{AB}$ to denote the full 6D
metric, Greek indices for the non-compact
four dimensions, and Roman letters for
components of tensors in the internal directions.

\section{The Unperturbed Model}
\label{sec:UnpFoot}

The authors of \cite{car} and \cite{navarro}
consider a six dimensional
model with two dimensions compact.
The bulk geometry has the topology of a sphere with metric
\beq
ds^2 = r^2 g_{ij} dx^i dx^j= r^2
[ d \theta ^2 +  \beta^2 \sin ^2 \theta d \alpha^2 ]
\eeq
and $\alpha$ has period $2\pi $.

The field content in the bulk is six -dimensional
gravity together with a $U(1)$ gauge field.
The field strength for the gauge field is non-vanishing
on the sphere,
\beq
F_{ij} = \sqrt{G^{(2)}} \epsilon_{ij}  B ~,
\eeq
with $\epsilon_{\theta \phi} \equiv 1$.
The corresponding flux
\beq
\Phi = \int d^2 x \sqrt{G^{(2)}} \epsilon^{ij} F_{ij}
\eeq
is conserved.

There is also a bulk cosmological constant $\Lambda$.
Obtaining a static solution requires
\beq
\Lambda = B^2/2
\eeq
which is the usual cosmological constant problem.
The radius of the sphere
is
\beq
r = {(\kappa B)}^{-1}
\eeq
where $\kappa^2$ is the six-dimensional Newton's constant.
What is intriguing about this model is that
if we add a brane at the north and south poles $(\theta=\theta^i)$,
each with tension $f^4$,
corresponding to a stress tensor
\beq
T_{\mu \nu} = - g_{\mu \nu} {f^4 \over 2 \pi}
{\sqrt{g^{(4)}} \over \sqrt{ G^{(6)}}}
\sum _i \delta(\theta -\theta^i) ~,
\eeq
a new static solution is obtained
without any additional fine tuning. The new geometry is still
locally a sphere,
but now there is a deficit angle
\beq
\gamma = \kappa^2 f^4
\label{delta-tension}
\eeq
corresponding to a local change in curvature at
the locations of the branes,
\beq
R_{ij} = g_{ij} + g_{ij} {\gamma \over \beta}
\sum_i {\delta(\theta-\theta^i) \over 2 \pi \sin \theta} ~.
\label{curv-int}
\eeq

In terms of the parameters above,
\beq
\beta = 1 - {\gamma \over 2 \pi} ~.
\label{beta-def}
\eeq
It is
transparent that the geometry is locally still a
sphere, since
we may rescale $\alpha$ so that
\beq
ds^2 =r^2[ d \theta^2 + \sin ^2 \theta d \phi^{ 2} ]
\label{backmetric}
\eeq
but now $0 \leq \phi  \leq 2 \pi \beta$.
This coordinate system is a physically
convenient choice for the linear
perturbation analysis, since all the effect
of the tension and deficit angle
is put into the boundary.

With a non-vanishing tension the area of
the sphere is smaller due to the deficit angle.
As a result, the magnetic flux
depends on the tension.
If a quantization condition is imposed on the flux, then
the self tuning feature of this model would no longer
work, for then
the cancellation of bulk parameters needed
to obtain a static solution would depend on the
tension \cite{monopole}.

To avoid this conclusion we will assume that there
is no quantization condition. This requires
us to assume that there are no electric sources
for the bulk $U(1)$ gauge field \footnote{This assumption 
does not affect the stability analysis.}. Since the
Standard Model fields are not charged under
this gauge group, this is not necessarily a
phenomenological problem.

\section{Linear Analysis}

\subsection{Gauge Fixing and Perturbed Equations}

A general perturbation of the background metric (\ref{backmetric})
is given by
\bea
ds^2 &=& (\eta_{\mu \nu} + h_{\mu \nu} ) dx^{\mu} dx^{\nu}
+ 2 h_{\mu \theta} dx^{\mu} d \theta + 2 h_{\mu \phi}
dx^{\mu} d \phi \nonumber \\
& & + (1+ h_{\theta \theta}) d \theta^2 +
2 h_{\theta \phi} d \theta d \phi
+ \sin^2  \theta (1 + \tilde{h}_{\phi \phi})
d \phi ^{ 2} ~.
\eea
Here we work in units with $r=1$.
For non-zero tension
there are curvature singularities
at $\theta =0$ and $\pi$. Then
$0 \leq \phi  \leq
2 \pi \beta$.
When the tension vanishes, there are coordinate singularities
at these points.

A convenient gauge choice would be
Gaussian Normal-like
gauge conditions in the bulk. These are
\beq
h_{\theta \mu} = h_{\theta \theta} = \partial^{\mu } h_{\mu \phi}
=0 ~,
\eeq
providing six conditions.

There is an important subtlety in choosing this
gauge though, which we now discuss. In the end we will
choose a gauge that is very similar to this one.

To set $h^{new}_{\theta \theta}=0$ one chooses a
gauge parameter
\beq
\epsilon_{\theta}(x, \theta) = -{1 \over 2}
\int^{\theta}_{0} d \theta^{\prime}
h^{old}_{\theta \theta}(x, \theta^{\prime}) ~.
\eeq
The problem with this gauge transformation
is with the location of the boundaries.
In the gauge with
$h_{\theta \theta}=0$, the brane at the north
pole is located at $\theta =0$, but the
brane at the south pole is now located at
$\overline{\theta} = \pi +  \epsilon_{\theta}(x, \pi)$, or
\beq
\overline{\theta} = \pi - F(x) \pi/2
\eeq
in general ($F$ is defined by these two equations).
Imposing
boundary conditions
at the location of
the south pole brane is technically too subtle
in this gauge.

Since $F$ represents a perturbation that cannot
be gauged away, it is more convenient to
put it in the metric rather than the location of
the boundary.
This is done
by choosing a slightly different gauge parameter,
\beq
\epsilon_{\theta}(x,\theta) = -{1 \over 2} \int^{\theta}_{0}
d \theta^{\prime}
h^{old}_{\theta \theta}(x,\theta^{\prime}) +  \theta F(x)/2 ~.
\eeq
which keeps the branes located at $0$ and $\pi$.
The price we pay is that we cannot completely gauge
$h_{\theta \theta}$ away, for in this gauge $h^{new}_{\theta
\theta}=F(x)$.

We use the $U(1)$ gauge invariance to set
\beq
\partial^{\mu} a_{\mu}=0 ~.
\label{gaugechoice1}
\eeq
This together with
\beq
h_{\theta \mu} = \partial^{\mu } h_{\mu \phi} =0
\label{gaugechoice2}
\eeq
and
\beq
h_{\theta \theta}(\theta, x)= F(x)
\label{gaugechoice3}
\eeq
represent our gauge conditions.

We search for massless scalar perturbations only,
since
only these
lead to
possibly dangerous long distance
deviations from Einstein gravity.

In addition, we focus on scalar perturbations
that are independent of the angular coordinate
$\phi$. The reason is that only $\phi$-independent
scalar perturbations couple to matter on the
brane, either through kinetic mixing with
the graviton or because of a non-vanishing
wavefunction at the brane locations. Perturbations with non-trivial
$\phi$ dependence only ``see'' the brane
tension through a change in their periodicity, and will
will therefore still vanish at the location of the branes.

All this
gauge fixing leaves seven scalar
perturbations,
\bea
ds^2 &= &\left(\eta_{\mu \nu}
+ \partial _{\mu} \partial _{\nu} \lambda +
{1 \over 4} \eta_{\mu \nu} h_{(4)} \right)
d x^{\mu} d x^{\nu} \nonumber \\
& & +(1+ F) d \theta^2 + 2 h_{\theta \phi} d \theta d \phi
+ \sin^2 \theta (1+ \tilde{h}_{\phi \phi}) d \phi ^{ 2}
\eea
and $A= a _{\theta} d \theta + a_{\phi} d \phi$
with field strengths $f_{AB} = \partial _A a_B - \partial_B a_A$.

Next Einstein's equations
\beq
E_{AB} = \kappa^2 T_{AB}
\eeq
in units with $r_0=1$ are obtained.
Focusing on $\phi$-independent perturbations,
the linearized Einstein equations in the bulk
and in the gauges
(\ref{gaugechoice1}-\ref{gaugechoice3}) are given by
\bea
(\mu \nu): ~~
 0 &= &-{1 \over 2} (\Box + \partial ^2_{\theta}
+ \cot \theta \partial _{\theta} )
 \left(h_{\mu \nu}- \eta_{\mu \nu} h_{(4)} \right)
 -{1 \over 2} \eta_{\mu \nu} \partial ^{\rho} \partial ^{\sigma}
h_{\rho \sigma}
\nonumber \\
& & + {1 \over 2} \eta_{\mu \nu} (\Box \tilde{h}_{\phi \phi}
+\Box F +F- \tilde{h}_{\phi \phi} )
+ {1 \over 2 }
\left( \partial _{\mu} \partial _{\rho}
h^{\rho} _{\nu}
+(\nu \leftrightarrow \mu)
\right)
\nonumber \\
& & +{1 \over 2} \eta_{\mu \nu} \left( \partial ^2 _\theta
\tilde{h}_{\phi \phi} +2 \cot \theta \partial _{\theta}
 \tilde{h}_{\phi \phi} \right)
\\ \nonumber
& &
 - {1 \over 2} \partial _{\mu} \partial _{ \nu} h_{(4)}
 - {1 \over 2} \partial _{\mu}
\partial _{ \nu} \tilde{h}_{\phi \phi}
- {1 \over 2} \partial _{\mu}
\partial _{ \nu} F
+ \eta _{\mu \nu}
{\kappa^2 B \over \sin \theta }
\partial _{\theta} a_{\phi}
\label{munueqn}
\eea
\bea
(\theta \mu): ~~
0 & = &
 \partial _{\theta} \partial ^{\rho} h_{\rho \mu}
+\partial_{\mu } \left( \cot \theta
(F-\tilde{h}_{\phi \phi})
-\partial_{\theta} h_{(4)} -\partial_{\theta}\tilde{h}_{\phi \phi}
-2 \kappa^2 B { a _{\phi} \over \sin \theta} \right)
\label{muthetaeqn}
\\
(\phi \mu): ~~
0 &=&  -\left( \Box+ \partial^2 _{\theta} -\cot \theta
\partial _{\theta} \right)
h_{\phi \mu} \nonumber \\
 & & + \partial _{\mu}
\left(
\partial_{\theta} h_{\theta \phi}
+ \cot \theta h_{\theta \phi}
+ 2 \kappa^2 B \sin \theta a_{\theta} \right)-2 \kappa^2 B
\sin \theta \partial_{\theta} a_{\mu}
\label{muphieqn} \\
(\theta \theta): ~~
0 & = & \Box h_{(4)}
- \partial^{\mu} \partial^{\nu} h_{\mu \nu}
+ \Box \tilde{h}_{\phi \phi} + \cot \theta \partial _{\theta}
h_{(4)}  + F
+ \tilde{h}_{\phi \phi}
\nonumber \\
& & - 2 \kappa^2 B { \partial_{\theta} a _{\phi} \over \sin \theta}
\label{thetathetaeqn} \\
(\theta \theta- \phi \phi): ~~
0 &=&    \partial^2 _{\theta} h_{(4)}
- \cot \theta \partial _{\theta} h_{(4)}
+ \Box F -\Box \tilde{h}_{\phi \phi}
\label{thetatheta-phiphieqn} \\
(\theta \phi ): ~~
0 &=&   -\Box h_{\theta \phi}  ~.
\label{thetaphieqn}
\eea
The linearized $U(1)$ gauge equations in these gauges
are
\bea
(\mu): ~~ 0 &=& \left( \Box_{(4)} + {\partial ^2 \over \partial \theta^2}
+ \cot \theta  { \partial \over \partial \theta}
\right) a_{\mu} - \partial_{\mu} (\nabla \cdot a)
-{ B \over \sin \theta }
 \partial _{\theta} h_{\mu \phi}
\label{amueqn}
\\
(\theta): ~~ 0 &=&
\Box_{(4)}
a_{\theta} \label{athetaeqn}
\\
(\phi): ~~ 0 &=& \sin \theta \partial _{\theta}
\left( {1 \over \sin \theta} \partial _\theta
a _{\phi} \right)  + \Box a_{\phi} - {B \over 2}\sin \theta
(\partial_{\theta}
\tilde{h}_{\phi \phi}  -\partial _{\theta} h_{(4)})
\label{aphieqn}
\eea
with $\nabla \cdot a = \partial_{\theta} a_{\theta}
+ \cot \theta a_{\theta}$ in the Lorentz gauge
and acting on $\phi$ independent perturbations,
and $\Box = \partial^{\mu} \partial_{\mu}$. For
massive states, (\ref{amueqn}) together
with the gauge choices (\ref{gaugechoice1})
and (\ref{gaugechoice2})
imply
$\nabla \cdot a =0$.
These equations in the bulk must be supplemented with
boundary conditions imposed at the locations of
the conical singularities. Boundary conditions
are discussed in the next section.

We emphasize that the equations obtained above are valid
for arbitrary brane tension. Due to the choice of parameterization
of the background given by (\ref{backmetric}), the effect of the brane
tension appears only in the boundary conditions,
and for solutions
with non-trivial $\phi$ dependence
(which we are not looking at here), in
$2 \pi \beta$ periodicity conditions.

As may be anticipated by considering the four-dimensional
effective potential \cite{sundrum1,effpot2}, the radion in
this model is massive. It is given by the mode
\beq
h_{\theta \theta}=\tilde{h}_{\phi \phi} = -{h_{(4)} \over 4} = F(x) ~,
\eeq
with all other fields vanishing. It has
a mass $m^2 =1/r^2_0$. This result agrees with
previous computations \cite{cline}.

In the remainder of this note we focus on massless modes only.

For the zero modes there is in addition the
usual residual gauge invariances $\Omega$ where
$\Box \Omega=0$. Here
we have a residual $U(1)$ gauge invariance
$\Lambda$, and residual diffeomorphism
invariances $\epsilon_{\phi}$ and
$\epsilon_{\mu}$
only, since $\epsilon_{\theta}$
is fixed by our gauge choice.
We use
$\Lambda$ and $\epsilon_{\phi}$  to set the
zero modes of
of $a_{\mu}$ and $h_{\phi \mu}$ to
be purely transverse, and
$\epsilon_{\mu}$ to set
the vector components of the four-dimensional
zero mode graviton to zero.

In total there are naively 10 equations for 7 variables.
But fortunately not all of these equations are independent.
Equation (\ref{aphieqn}) and the trace of (\ref{munueqn})
are derivable from
other equations. Equations (\ref{thetaphieqn})
and (\ref{athetaeqn}) are
trivial acting on the zero modes we
are focusing on.
This leaves
six non-trivial equations for seven
scalar perturbations (but recall that $F$ is pure gauge
in the bulk).

Inspecting the equations further, $a_{\theta}$ and
$h_{\theta \phi}$ decouple from the other perturbations.
Their wavefunctions are
fixed by (\ref{muphieqn}) and
(\ref{amueqn}) to be
\beq
a_{\theta} = {c_0(x) \over \sin \theta} ~~~,~~~ h_{\theta \phi}
=2 \kappa^2 B c_0(x) \cot \theta  ~.
\end{equation}

Next we proceed to solving the other equations.
Equation (\ref{thetatheta-phiphieqn})
can be solved immediately to give
\beq
h_{(4)} = c_1(x) - c_2(x) \cos \theta  ~.
\label{h4sol}
\end{equation}
Next use (\ref{thetathetaeqn}) to
solve for $\tilde{h}_{\phi \phi}$ and
substitute this into  (\ref{muthetaeqn})
to obtain
\bea
{\partial^2 a_{\phi} \over \partial \theta ^2 }
 + a_{\phi} = -{c_2(x) \over 8 \kappa^2 B}
\left(7 -11\cos^2 \theta\right)
+{F(x ) \over \kappa^2 B} \cos \theta
\eea
This will have two homogeneous solutions and
one homogeneous solution. Since the sources
are independent, we may think of this as
four solutions in total.
Finally, (\ref{munueqn})
with $\mu \neq \nu$ determines
$\lambda$ in terms of the previous solutions.

The most general solutions to
these equations are given by
\bea
\tilde{h}_{\phi \phi} &=&c_3(x) + {5 \over 6} c_2(x) \cos \theta
+2 \kappa^2 B c_4(x) \cot \theta
+ \theta \cot \theta F(x)  \nonumber \\
a_{\phi} &=&-{c_3(x) \over 2 \kappa^2 B} \cos \theta
+{c_2(x)\over 24 \kappa^2 B}(1 - 11 \cos^2 \theta)
 + c_4(x) \sin \theta  \nonumber \\
& & +{F(x) \over 2 \kappa^2 B} \theta \sin \theta
\nonumber \\
h_{\theta \theta} &=& F(x)
\eea
together with (\ref{h4sol}).

To summarize, for an arbitrary tension
we have found the
the most general
$\phi-$independent
massless scalar perturbation
solution to the $U(1)$ and
Einstein field equations.
Imposing boundary conditions on these solutions
is discussed in the next section.

\subsection{Boundary Conditions}

For non-zero tension the locations of the
branes are special points on the sphere with
curvature singularities.
To obtain the boundary conditions for the perturbations
at these points
we need to inspect the field equations
and match the singularities.

It is useful to rewrite the internal metric,
including perturbations, as
\beq
ds^2_{int} = r^2_0 \left(1 + h_{\theta \theta} \right)
\left[ d \theta^2 + \beta^2 \sin ^2 \theta (1 +
\tilde{h}_{\phi \phi}- h_{\theta \theta} ) d \alpha^2 \right] ~,
\eeq
where in a change of notation, $r_0= (\kappa B)^{-1}$ denotes
the unperturbed
radius.
To linear order in the perturbations this is
equivalent to our previous parameterization.
Here we are also only focusing on $h_{\theta \theta}$ and
$\tilde{h}_{\phi \phi}$, since $h_{\theta \phi}$ decouples
and does not couple to brane matter.
For long-wavelength perturbations this describes
a new background with effective radius
\bea
r^2 &=& r^2_0 (1 + h_{\theta \theta})
\label{r-eff}
\eea
and effective deficit angle
\bea
\tilde{\beta} ^2 &=& \beta^2 ( 1 + \tilde{h}_{\phi \phi} - h_{\theta
\theta}) ~.
\label{beta-eff}
\eea

To obtain first order
terms in the equations of motion with
explicit delta-function singularities,
which correspond to
perturbations without derivatives,
we may use
(\ref{curv-int}), using (\ref{beta-def})
and substituting for the effective radius
(\ref{r-eff}) and effective deficit angle (\ref{beta-eff}).

This implies that to linear order there are no singular
contributions to the $(i,j)$ Einstein equations.
There is however, a singular contribution to the
$(\mu, \nu)$ equation, given by
\bea
E_{\mu \nu} |_{sing.} & = & \sum_i g_{\mu \nu} {1 \over \tilde{\beta} r^2}
 \left[\tilde{\beta}-1 \right]
{\delta(\theta -\theta^i) \over \sin \theta} \nonumber \\
&=& \kappa^2 T_{\mu \nu} |_{sing.} \nonumber \\
&=& -\kappa^2 f^4  \sum_i g_{\mu \nu} {\sqrt{g^{(4)}} \over
2 \pi \sqrt{G^{(6)}} }
{\delta(\theta - \theta^i) }
= -\kappa^2 f^4  \sum_i {g_{\mu \nu} }
 {1 \over \tilde{\beta} r^2}
{\delta(\theta - \theta^i) \over 2 \pi \sin \theta}
\label{sing-uv}
\eea
Before concluding that these singularities
must match, we must check that perturbations involving
derivatives
cannot contribute
additional singularities. To see that they cannot,
use the $(\mu \theta ) $ equation to solve
for $\tilde{h}_{\phi \phi}$ and substitute it into
the $(\mu = \nu)$ equation. One finds all
the terms cancel identically, except for
the singularities appearing in (\ref{sing-uv}).
Inspecting (\ref{sing-uv}), we see that
to satisfy the equations of motion requires
\beq
[\tilde{\beta}-1]| = -{\kappa^2 \over 2 \pi} f^4
\eeq
or
\beq
[{\tilde{h}}_{\phi \phi} - h_{\theta \theta} ]| = 0
\label{bc2}
\eeq
at the location of either brane.
This is just the
statement that without a change in the
tension, a perturbation in
the metric cannot change the deficit
angle.

To obtain other boundary conditions we must inspect
the other field equations.

The $(\theta \theta - \phi \phi)$ metric equation
does not contain any curvature singularities. Requiring
that the solution is finite gives the boundary condition
\beq
\partial _{\theta} h_{(4)} | =0 ~.
\label{bc1}
\eeq

Finally, to obtain the boundary condition for
$a_{\theta}$,
consider spreading the brane out into a ring located
at $\theta \sim \epsilon$.
Later we will send $\epsilon \rightarrow 0$. Inside
the ring we have no deficit angle, so all fields
are regular at the pole, and therefore $a_{\theta} \rightarrow 0$.
The ring does not affect the
$a_{\mu}$ gauge equation of motion, so $\nabla \cdot a =0$ is
the $a_{\theta}$ equation both inside and outside the ring.
The only solution inside the ring, consistent with this
equation and the boundary condition at the pole, is
$a_{\theta }=0$. By continuity at the location of
the ring, the boundary condition outside the ring is then
\beq
a_{\theta} | =0 ~.
\label{atheta-bc}
\eeq
Equations (\ref{bc2}), (\ref{bc1}) and (\ref{atheta-bc})
are our boundary conditions when the
tension is non-vanishing.

For vanishing tension, there are stronger
constraints than these from requiring
that the north and south poles not be special
points. This means that our solutions in polar coordinates
should have sensible ({\cal C}) values when expressed in a
Cartesian basis local to the poles.
This implies
that near a pole,
\beq
h_{\mu \phi} \sim f_{\mu \phi} \sim
\tilde{h}_{\phi \phi} -h_{\theta \theta} \sim \theta^2
\eeq

Inspecting the four solutions above, we first note
that the normal mode $c_0(x)$ and $c_4(x)$ are too singular to
satisfy the boundary conditions.
However the linear combination
\beq
c_3(x)=-{5 \over 6} c_2(x)
\eeq
and the normal mode $F(x)$ both independently
satisfy the boundary conditions at $\theta =0$.
Thus there are two independent solutions
that satisfy the boundary conditions at
$\theta =0$.

But neither solution satisfies the boundary
conditions at $\theta =\pi$. As $\theta \rightarrow \pi$, a
general combination of these two solutions behaves as
\bea
\tilde{h}_{\phi \phi}-h_{\theta \theta}
& \rightarrow & -{5 \over 3}
c_2(x)- F(x) - {\pi \over \sin \theta} F(x) ~.
\eea
The boundary condition
implies that the left side should vanish at
the boundary, so
both $c_2=F$=0. Thus there are no solutions that
satisfy the boundary conditions at both $\theta =0$
and $\theta =\pi$.

This conclusion is true
for arbitrary tension. It is straightforward
to repeat the analysis for vanishing tension.
The only difference is that the boundary
conditions are {\em stronger}, because
of the constraint of regularity. No zero
modes therefore exist in this limit either.
Thus
any light mode with mass that vanishes
as the tension is sent to zero is also excluded
by our analysis.

\section{Fine-tuning or Self-tuning?}

We return to an issue briefly issue raised in the
introduction. One might think that
the relation (\ref{delta-tension}) between the deficit angle and
the brane tension does not represent a fine-tuning, since
the deficit angle is an integration parameter, not
a parameter of the Lagrangian. But a simple
four-dimensional counter-example illustrates that the issue is
not as straightforward \cite{weinberg}.

Consider a
four dimensional theory with a bare cosmological
constant $\Lambda_0$ and
a four-form field strength with value
\beq
F_{\mu \nu \rho \sigma} = c \epsilon_{\mu \nu \rho \sigma}
\eeq
which satisfies the field equations of motion and
where $c$ is an integration parameter \cite{fourform}. The source for
gravity in this theory is
\beq
\Lambda= \Lambda_0 + c^2 ~.
\eeq
The integration parameter may be chosen
to be
$c^2 = -\Lambda_0$, giving a flat space solution.
But obviously this is not the only solution,
as there is a family of
de-Sitter and Anti-de-Sitter solutions.

In analogy with this four-form example, does
setting the deficit angle - an integration parameter -
of the six-dimensional model
to be equal to the tension involve a fine-tuning? 
For recall that it has not been demonstrated that the 
deficit angle is forced by the equations of 
motion to be equal to the tension--that only followed 
from the equations of motion after assuming a flat-space 
ansatz. For maximally symmetric space-times, 
\cite{cline} has found 
that once the bulk cosmological 
constant is finetuned 
against the magnetic flux, then de-Sitter or 
Anti-de Sitter solutions along the brane directions 
are forbidden, and the deficit angle is equal to the tension. 

But that still does not completely 
address the issue. In other words, 
it isn't clear that a change in the tension is canceled 
by a change in the deficit angle. If other cosmological 
solutions exist, for the same tension but other values of 
the deficit angle, then (\ref{delta-tension})
is a fine-tuning, and this model would then be less appealing.

What might these solutions look like? \footnote{The authors 
thank Maxim Perelstein for discussions on this point.} 
A dynamical change in the tension could lead to a 
solution that is not maximally
symmetric, but instead interpolates 
between an  
inflating or generally time-dependent solution near 
the brane, to a static geometry far from the brane.
Far from the brane, the deficit angle would be (approximately) 
equal  
to its unperturbed value. 
 


\section{Conclusions}

We have performed a linear perturbation analysis
of the model presented in \cite{car} and \cite{navarro}
to search for phenomenologically
dangerous massless or approximately massless
scalars. After imposing
the boundary conditions, we have found
that there are no such modes. 

If this model does have a self-tuning mechanism, then
our results do raise a puzzle. Namely, below
the compactification scale the only light modes
are the four dimensional graviton, the Standard Model
fields, a gravi-vector boson from the residual isometry of the bulk,
and a $U(1)$ gauge
boson. The latter two do not couple to the tension on the
brane and so we can forget about them.
The puzzle is that one might have expected 
that the self-tuning of a small enough change 
in the brane tension could be understood 
in the four dimensional effective 
theory--but this does not seem likely, 
since our results show that 
this model lacks any additional light degree 
of freedom.

If this is the case here,
then in order for this model to be compatible with the observed
size or bound on the cosmological constant,
we would need $1/r \sim 10^{-3}$ eV \cite{luty}.
There is an additional puzzle here, if 
the model does self-tune: from the higher dimensional perspective,
it appears that it is the full quantum mechanical
tension on the brane that is canceled, not just
the high energy contribution. If this were true,
then there would not be a constraint on the
size of the internal space arising from 
vacuum fluctuations of brane localised matter.

Our results suggest that 
if there is self-tuning, then it must be 
due to modes no lighter 
than the compactification scale. 
A consistent story would then be that 
the massive states only partially cancel a 
dynamical change in the tension, down to an amount 
set by the compactification scale. 

Finally, demonstrating that these models 
do or do not self-tune might be difficult: one would 
need to find or demonstrate the absence of 
non-maximally symmetric 
solutions with an approximately static 
deficit angle and 
geometry far from the brane, but with a cosmological geometry near 
the brane. 
That is, to follow the cosmological 
evolution of this system through a phase transition.

\section*{Acknowledgments}

We would like to thank K. Akama, 
C. Burgess, C. Csaki, 
M. Giovannini, C. Grojean, M. Luty, M. Perelstein, J. Vinet 
and M. Wise
for discussions and comments. 
This worked is supported by the U.S.
Department of Energy under contract number
DE-FG-0392-ER40701.


\end{document}